\documentclass[11pt]{article}
\usepackage{lscape}
\usepackage{graphicx}
\graphicspath{ {./images/} }
\usepackage[utf8]{inputenc}
\usepackage[titletoc]{appendix}
\usepackage[usestackEOL]{stackengine}
\usepackage{lipsum}
\edef\tmp{\the\baselineskip}
\setstackgap{L}{\tmp}
\usepackage{amsmath,amssymb,tabularx,booktabs}
\usepackage{extarrows} 
\usepackage{float}
\begin{document}
\title{
Importance of strange sea to the charge radii and quadrupole moment of $ J^P=\frac{1}{2}^+,\frac{3}{2}^+$  baryons}
\author{Preeti Bhall, A. Upadhyay}
\date{School of Physics and Material Science, \\
Thapar Institute of Engineering and Technology, Patiala, Punjab-147004\\
\today}
\maketitle
\begin{abstract}
A statistical framework in conjugation with the principle of detailed balance is employed to examine the low-energy properties i.e. charge radii and quadrupole moment of J$^P$=$\frac{1}{2}^+$ octet and J$^P$=$\frac{3}{2}^+$ decuplet baryon. The statistical approach assumes the expansion of baryonic system in terms of quark-gluon Fock states. We systematically apply operator formalism along with the statistical approach to study the charge radii and quadrupole moment of baryons. Based on the probabilities of all possible Fock states in spin, flavor and color space, the importance of sea with quarks and gluons is studied. 
The individual contribution of the constituent quarks and sea which contains terms of scalar, vector and tensor is explored. Due to large mass difference between strange and non-strange content, the SU(3) breaking effect are also investigated. The extent to which strange quark-antiquark pair is considered in sea is constrained by the mass of hadrons, the free energy of gluons in conformity with experimental indications. We focus on the individual contribution of strange and non-strange sea (g, $\langle u\bar u\rangle$, $\langle d\bar d\rangle$ and $\langle s\bar s\rangle$) accomodability in the respective hadrons for their charge radii and quadrupole moment. The present work has been compared with various theoretical approaches and some known experimental observations. Our computed results may provide important information for upcoming experimental findings.
\end{abstract}
\section{Introduction}
Recently, a phenomenal discovery in the field of particle physics was celebrated and honored by three Nobel Prizes. Scientists Pierre Agostini, Ferenc Krausz, and Anne L'Huillier were awarded with Nobel Prize in October 2023 "for experimental methods that generate attosecond pulses of light for the study of electron dynamics in matter".  
Each advancement motivates physicists to understand better the internal structure of ordinary matter. The low energy properties (masses, spin distribution, magnetic moment, semi-leptonic decays, charge radii etc.) are preferred to study as they reveal the distribution of quarks inside the hadrons. The electromagnetic form factors 
are the most basic quantities 
that provide essential information about the static properties of hadrons. 
A lot of progress has been observed in both experimental and theoretical approaches for the study of the intrinsic structure of hadrons. Recently in 2022, the new AMBER experiment at the CERN SPS \cite{01} examined the proton and meson charge radii, the mesonic Parton momentum distributions, the kaon polarizability, and the kaon-induced hadron spectroscopy. 
In 2021, at the Beijing Spectrometer (BESIII), form factors are measured for different baryons in the time-like region \cite{02}. We know that the hadrons, in addition to the valence quarks, also carries  dynamic "sea" which is filled with different flavors of quark-antiquark pairs ($u\bar u, d\bar d, s\bar s$) and gluons. 
The importance of sea is evident from the fact that 30$\%$ of the total nucleonic spin is carried by quark–antiquark pairs present in the sea. 
In 2019, the STAR experiment at the Relativistic Heavy Ion Collider (RHIC) observed the major contribution of sea antiquarks to the spin distribution of proton \cite{03}.
Moreover, the strange quark is one of the essential components of intrinsic sea that significantly contributes to spin distribution among quarks and gluons within nuclei \cite{04, 05}. 
The NuTeV collaboration \cite{06} at FermiLab predicted the non-zero value of the quark contribution to the spin of nucleon via the strange-quark content ratio. The ratio is the fraction of nucleon momentum that is carried by strange quarks to that carried by non-strange quarks $\frac{2(s+ \bar s)}{u+ \bar u+ d+ \bar d}$ = 0.477$\pm$0.063$\pm$ 0.053 \cite{07}. This ratio implies the existence of strange quarks in the sea. Several experiments like  SAMPLE at MIT-Bates \cite{08}, HAPPEX at JLab \cite{09,10,11}, and PVA4 at MAMI \cite{12} measured the weak and electromagnetic form factors from the elastic scattering which describe the significance of strange quarks in the charge, current, and spin structure of the nucleon. The complete information about strange quark effects in the hadronic sea is still being explored. The upcoming experimental facility at FAIR, $\bar P$ANDA-GSI is expected to study hyperons, particularly at low energy regime \cite{13,14} as well as a part of BESIII experiment shall include the strange quark systems \cite{15} and J-PARC facility \cite{16}. \\
The electromagnetic properties of baryons i.e. charge radii and quadrupole moment yield important information about internal structure within the non-perturbative regime of QCD (Quantum Chromodynamics). They describe the spatial charge distribution inside the baryons and provide information about their geometrical size and shape. The experimental predictions of charge radii of octet particles i.e. Proton ( $r_p$ = 0.8409 $\pm$ 0.0004 fm), Neutron ( $r^{2}_n$ = -0.115 $\pm$ 0.0017 fm$^2$),  Sigma ($\Sigma^-$= 0.78 $\pm$ 0.10 fm) listed in Particle Data Group (PDG) \cite{17}. For J$^P$ = $\frac{3}{2}^+$ particles, the experimental data are very limited due to their shorter lifetime. In literature, various theoretical approaches such as non-relativistic quark model \cite{18}, 
relativistic quark model \cite{19}, light-front holographic method \cite{20}, QCD sum rule \cite{21, 22}, 
1/N$_c$ expansion method \cite{23, 24}, Lattice QCD \cite{25,26}, chiral constituent quark model ($\chi$CQM) \cite{27,28}, General Parameterization (GP) method \cite{29,30} have studied the several properties of octet and decuplet baryons. 
In Ref. \cite{20}, the authors computed the form factors, masses, magnetic moment, and charge radii of octet and decuplet baryons. Using the QCD sum rule approach in ref. \cite{21,22}, the author estimated the magnetic dipole, electric quadrupole, and magnetic octupole moments of $\Delta$-baryons. A.J. Buchmann and E.M. Henley \cite{29,30} calculated the quadrupole moment of nucleons and decuplet baryons using Morpurgo’s general QCD parametrization method. Based on the framework of chiral constituent quark model ($\chi$CQM), N. Sharma and H. Dahiya \cite{27,28} studied the effect of SU(3) symmetry and its breaking to the charge radii and quadrupole moment of baryons. In ref. \cite{25, 26}, the charge radii of decuplet baryons have been studied in lattice QCD. 
Despite the significant success in experimental and theoretical domains, 
the information on the structure of baryons are still not well understood.\\ 
For the purpose of better understanding, we study the charge radii and quadrupole moment of J$^P$=$\frac{1}{2}^+$ octet and J$^P$=$\frac{3}{2}^+$ decuplet baryons in the framework of statistical model with the principle of detailed balance. This model assumes the hadrons as a complete set of quark–gluon Fock states. The principle of detailed balance is associated with the probability of finding different Fock states inside the hadrons. Our aim is to calculate individual contributions from strange as well as non-strange components inside octet and decuplet baryons for the above-mentioned properties. To study SU(3) breaking in valence and sea, we need to incorporate a parameter that studied the effects of strange quark mass on the charge radii and quadrupole moment. The probabilities associated with various Fock states are affected by considering the strange quark in the sea. With the inclusion of 'sea quarks' various low-energy properties like masses [31], spin distribution [32], magnetic moments [33], the importance of sea contribution to nucleons [34] are successfully analyzed using statistical model. The statistical approach provides a strong base for a deep understanding of baryonic structure. \\ 
This paper is organized in the following manner: Sec. II includes a detailed discussion of the wavefunction of octet and decuplet baryons with sea components. In Sec. III, the operator formalism is briefly discussed. 
Sec. IV, describes the principle of detailed balance and the statistical approach for $ J^P=\frac{1}{2}^+,\frac{3}{2}^+$ baryons having various quark-gluon Fock states including strange quark-antiquark pairs. Numerical results of charge radii and quadrupole moment with SU(3) symmetry and its breaking are presented in section IV. Concluding remarks are given in section V.


\section{Theoretical Formalism}
Hadrons are, in the first approximation, composed of constituent quarks like baryons (qqq) and mesons ($q \bar q$), with appropriate spin-flavor and color singlet combinations. In Quantum Chromodynamics (QCD), the presence of quark-gluon interaction suggests that a hadron can be understood as having valence quarks, enveloped by a surrounding "sea" that contains infinite no. of virtual quark-antiquark ($q \bar q$) pairs multiconnected through gluons. However, various studies have shown that the sea contribution may change the structure of hadrons and modify the low-energy properties. The "sea" is characterized by its quantum numbers such as flavor, spin, and color. The quantum numbers are chosen in a way that ensures the combined effects of sea and valence quarks yield the intended quantum numbers for the observed baryon. The valence quark wavefunction of baryon \cite{35} can be written as:\\
\begin{equation}
    \Psi= \Phi (|\phi\rangle |\chi\rangle |\psi\rangle |\xi\rangle)
\end{equation}
where $|\phi>$, $|\chi>$, $|\psi>$ and $|\xi>$ denote the flavor, spin, color and space $q^{3}$ wave functions respectively. The valence quarks are considered to be in S-wave state so
that the space wavefunction $|\xi>$  is symmetric under the permutation of any two quarks. The color wavefunction $|\psi>$ is totally antisymmetric to give a color singlet baryon. Consequently, the flavor-spin wavefunction ( $|\phi>$, $|\chi>$) is totally symmetric to ensure the overall anti-symmetrization of the baryonic wavefunction. On the other side, sea is assumed to be flavorless but has appropriate spin and color wavefunction. For example- If we consider two gluons present in the sea, each having spin 1 and color octet '8', the spin and color space will yield the following  possibilities:
$$\mathbf{Spin}: gg: 1 \bigotimes 1 = 0_s \bigoplus 1_a \bigoplus 2_s$$
$$\mathbf{Color}: gg: 8 \bigotimes 8 = 1_s \bigoplus 8_s \bigoplus 8_a \bigoplus {10}_a \bigoplus \bar{10}_a \bigoplus 27_s$$
Here, the subscript s and a represent the symmetric and anti-symmetric combinations of the states, respectively. The total flavor-spin-color wavefunction of a spin up ($\uparrow$) baryon which consists three valence quarks with sea components can be written systematically as:\\
For octet baryons:\\
\begin{equation}\begin{aligned} |\Phi_{1/2}^{(\uparrow)}> = & \frac{1}{N} [{\Phi_{1}}^{(\frac{1}{2} \uparrow)}H_{0}G_{1} + a_{8} ({\Phi_{8}}^{(\frac{1}{2})}\otimes H_0)^{\uparrow}G_8 + a_{10} {\Phi_{10}}^{(\frac{1}{2}\uparrow)}H_{0}G_{\bar {10}} +\\& b_1({\Phi_1}^{(\frac{1}{2})} \otimes H_1)^{\uparrow} G_1+b_8 ({\Phi_8}^{(\frac{1}{2})} \otimes H_1) ^{\uparrow} G_8 + b_{10} ({\Phi_{10}}^{(\frac{1}{2})} \otimes H_1)^{\uparrow} G_{\bar {10}}\\& + c_8 ({\Phi_8}^{(\frac{3}{2})} \otimes H_1)^{\uparrow} G_8 + d_8 ({\Phi_8}^{(\frac{3}{2})} \otimes H_2)^{\uparrow} G_8\end{aligned}\end{equation}
where $N^{2} = 1+ a_{8}^{2} +a_{10}^{2} +b_{1}^{2} +b_{8}^{2} + b_{10}^{2} + c_{8}^{2} + d_{8}^{2}$\\
For decuplet baryons \cite{33}:\\
\begin{equation}
    \begin{aligned}
  |\Phi_{3/2}^{(\uparrow)}> =& \frac{1}{N} [a_0{\Phi_{1}}^{(\frac{3}{2} \uparrow)}H_{0}G_{1} + b_1({\Phi_1}^{(\frac{3}{2})} \otimes H_1)^{\uparrow} G_1 + b_8 ({\Phi_8}^{(\frac{1}{2})} \otimes H_1) ^{\uparrow} G_8 +\\&d_1 ({\Phi_1}^{(\frac{3}{2})} \otimes H_2)^{\uparrow} G_1 + d_8 ({\Phi_8}^{(\frac{1}{2})} \otimes H_2)^{\uparrow} G_8]   
    \end{aligned}
\end{equation}
where $N^{2} = a_{0}^{2} +b_{1}^{2} +b_{8}^{2} + d_{1}^{2} + d_{8}^{2}$\\
Here, N represents the normalization constant. The wavefunction consists various combination of both valence and sea part such as ${\Phi_{1}}^{(\frac{1}{2} \uparrow)}H_{0}G_{1}, ({\Phi_{8}}^{(\frac{1}{2})}\otimes H_0)^{\uparrow}G_8, {\Phi_{10}}^{(\frac{1}{2}\uparrow)}H_{0}G_{\bar {10}}$ etc. For example: the first term of octet wavefunction $\Phi_1^{(\frac{1}{2})}$ contains the valence spin- $\frac{1}{2}$, flavor octet and color singlet state along with sea component having spin-0 (H$_0$)and singlet color (G$_1$) state:\\
\begin{equation}
    \Phi_1^{(\frac{1}{2})}H_0G_1= \Phi (8,\frac{1}{2},1_c)H_0G_1
\end{equation}
The term $b_1(\Phi_1^{(\frac{1}{2})} \otimes H_1)^\uparrow G_1$ comes from vector sea (spin-1) combined with the spin- $\frac{1}{2}$ of core baryon and written as:\\
\begin{equation}
    b_1(\Phi_1^{(\frac{1}{2})} \otimes H_1)^\uparrow G_1 = \sqrt{\frac{2}{3}} B(8, \frac{1}{2}\downarrow)H_{1,1} - \sqrt{\frac{1}{3}} B(8, \frac{1}{2}\uparrow)H_{1,0}
\end{equation}
Similarly, all the terms of the octet and decuplet wavefunctions $b_1(\Phi_1^{(\frac{3}{2})} \otimes H_1)^\uparrow$, $b_1(\Phi_1^{(\frac{1}{2})} \otimes H_1)^\uparrow G_1 $, $b_8(\Phi_1^{(\frac{1}{2})} \otimes H_1)^\uparrow G_8 $, $d_1(\Phi_1^{(\frac{3}{2})} \otimes H_2)^\uparrow G_1$  etc. written with suitable C.G. coefficients by considering the symmetry property of the component wave function [].  The spin and color wavefunction of sea is specified by H$_{0,1,2}$ and G$_{1,8,10}$. The spin 0 (spin 1, spin 2) sea refers to the scalar (vector, tensor) sea. The coefficients ($a_0, a_8, a_{10}, b_1, b_8, b_{10}, c_8, d_8$) appear in eq. 2 and 3 are the statistical parameters that contain the contribution of scalar, vector and tensor sea. For octet baryonic wavefunction, the parameters $a_0, a_8, a_{10}$ give contribution of scalar sea i.e. spin 0. Similarly, the parameters $ b_1, b_8, b_{10}, c_8$ give the vector sea (spin 1) contribution and $d_8$ is signifying tensor sea (spin 2) contribution. These statistical parameters play an important role in computing the low-energy properties (spin distributions, masses, semi-leptonic decays) of baryons. The details of all the terms of the above-mentioned wavefunction can be found in refs. \cite{35,36}.
\section{Charge radii and quadrupole moment}
The internal electric and magnetic structure of hadrons is characterized by their electromagnetic form factors (FFs). They describe the geometrical shape, spatial distributions of electric charge and current within the baryons and thus are intimately related to their intrinsic structure. Most of our experimental knowledge on electromagnetic structure comes from the 
elastic and inelastic scattering experiments (eN $\rightarrow e' \Delta$, $\gamma N \rightarrow \Delta$ transition) facilitated at various laboratories like Jefferson Lab, MIT-Bates, ELSA, MAMI etc. \cite{37,38}. It has been proposed that the $\Delta$(1232) resonance - lowest lying excited state of the nucleon N(939) is closely related to the quadrupole deformation of the nucleon's ground state charge distribution \cite{39, 40,41}. The N$\rightarrow \Delta $  excitation (in short $\gamma N \Delta$) is allowed the magnetic dipole (M1), electric quadrupole (E2), and charge (or Coulomb) quadrupole (C2) transition \cite{42, 43} modes due to the parity invariance and angular momentum conservation as shown in fig.1. This excitation is determined in terms of three electromagnetic transition form factors $G^{ N\rightarrow \Delta}_{M1}$($Q^2$),  $G^{ N\rightarrow \Delta}_{E2}$($Q^2$) and  $G^{ N\rightarrow \Delta}_{C2}$($Q^2$).
\begin{figure}
    \centering
    \includegraphics{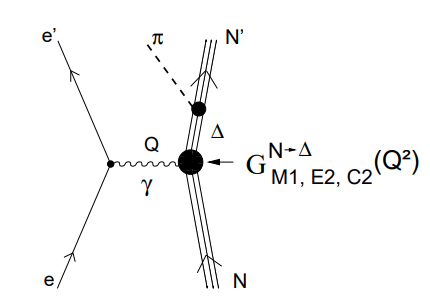}
    \caption{The exciation of $\Delta$-resonance is shown with inelastic electron-neutron scattering (e N$\rightarrow e' \Delta$) which is characterized by three electromagnetic transition form factors}
    \label{fig:(a)}
\end{figure}
Electric quadrupole (E2) matrix elements, including quadrupole moments, provide a principal measure of nuclear deformation, rotation, and related collective structure. The nonzero value of the E2 and C2 multipoles indicates that the charge distribution within the nucleon is not spherically symmetric, but it has an angular dependency $\rho(r)=\rho(r, \theta, \Phi)$ \cite{44,45}. If the charge distribution of the initial and final three-quark states were spherically symmetric, the value of E2 and C2 amplitudes of the multipole expansion would be zero. Many experiments observed that at low momentum transfer, the magnetic dipole (M1) amplitude dominates in N (939)$\rightarrow \Delta (1232) $ excitation. The amplitude (M1) involves the spin and isospin flip of a single quark. The ratio of the electric quadrupole amplitude to the magnetic dipole amplitude $\frac{E2}{M1} = -0.025\pm 0.005$ and the charge quadrupole amplitude to the magnetic dipole amplitude i.e. $\frac{C2}{M1}$ has been measured \cite{17,46,47}. 
The nucleonic vertex function in terms of Electromagnetic Dirac and Pauli form factors $F_{1}(Q^{2})$ and $F_{2}(Q^{2})$ is represented as:
\begin{equation}
\Gamma^{\mu}=F_{1}(Q^{2})\gamma^{\mu}+\kappa F_{2}(Q^{2}(i \frac{\sigma^{\mu\nu}q_{\nu}}{2m}))
\end{equation}
where $\kappa$ represent the anomalous part of the magnetic moment, $\gamma^{\mu}$ are Dirac matrices and $\sigma^{\mu\nu} = i(\gamma_\mu\gamma_\nu- \gamma_\nu\gamma_\mu)/2$. Further, the electric and magnetic Sach form factors  $G_{E}(Q^{2})$ and $G_{M}(Q^{2})$ can be related as \cite{48}:
\begin{equation}
G_{E}= F_{1}- \tau\kappa F_{2}
\end{equation}
\begin{equation}
G_{M}=F_{1}+\kappa F_{2}    
\end{equation}
where $\tau =(\frac{Q}{2m})^{2}$. The Fourier transform of the elastic form factors provide
insights about the radial variation of the charge $\rho$(r) and current j(r) densities.  The lowest moments of the charge density operator $\rho$ in the low-momentum expansion are charge radii ($r^2_{B}$) and quadrupole moment ($Q_{B}$). The general form of multipole expansion of the charge density operator $\rho$ up to $q^2$ limit is expressed as:\\
\begin{equation}
    \rho(q) = e- \frac{q^2}{6}r^2_{B}-\frac{q^2}{6}Q_{B}+...
\end{equation}
Here $q^2$ is the four-momentum transfer of the virtual photon. The first two terms come from the spherically symmetric monopole part and the third term is obtained from the quadrupole part of the charge density $\rho$. They characterize the total charge (e), spatial extension ($r^2_{B}$) and shape ($Q_B$) of the system. The mean square charge radii $r_{B}^2$ under spatial rotation is defines as:\\
\begin{equation}
    \langle r_{B}^2\rangle = \int d^{3}r \rho{(r)r^{2}}
\end{equation}
where $\rho(r)$ is the charge density. The deformity in the shape of baryons can be determined from the intrinsic quadrupole moment \cite{49,50}:\\
\begin{equation}
    Q_{0}'=\int d^{3}r \rho{(r)(3z^{2}-r^{2})}
\end{equation}
 If the charge
density is concentrated along the z-direction (symmetry axis of the particle), the term
proportional to 3$z^2$ dominates which makes $Q'_0$  positive, and the shape of the particle is prolate. If the charge density is concentrated in the equatorial plane perpendicular to z, the term proportional to $r^2$ prevails, $Q'_0$ is negative, and the shape of the particle is oblate.
In order to find the quadrupole moment of baryon, a general QCD unitary operator and QCD eigen states $|B\rangle$ are defined explicitly in terms of gluons and quarks. The quadrupole moment operator in terms of spin-flavor space can be expressed as \cite{51,52}:
\begin{equation}
 { \widehat{Q}_B = B\sum_{i\neq j}^3 e_i(3\sigma_{iz}\sigma_{jz}-\sigma_i.\sigma_j) + C\sum_{i\neq j\neq k}^ 3 e_i(3\sigma_{jz}\sigma_{kz}-\sigma_j.\sigma_k)} 
\end{equation}
Here, $\sigma_{iz}$ represents the z-component of the Pauli spin (isospin) matrix $\sigma_i$ and $e_i$ is the charge of the $i^{th}$ quark where i = (u,d,s). The expanded form of the quadrupole moment operator of octet and decuplet baryons \cite{27} can be written as:\\
\begin{equation}
    \widehat{Q}_{1/2}= 3B\sum_{i\neq j} e_i\sigma_{iz}\sigma_{jz}+ 3C\sum_{i\neq j\neq k} e_i\sigma_{jz}\sigma_{kz}+(-3B+3C)\sum_{i} e_i \sigma_{iz}+ 3B\sum_{i} e_i
\end{equation}
\begin{equation}
     \widehat{Q}_{3/2}= 3B\sum_{i\neq j} e_i\sigma_{iz}\sigma_{jz}+ 3C\sum_{i\neq j\neq k} e_i\sigma_{jz}\sigma_{kz}+(-5B+5C)\sum_{i} e_i \sigma_{iz}+ (3B-6C)\sum_{i} e_i
\end{equation}
Also, the expansion of charge radii operator \cite{24} can be written in terms of the sum of one-, two-, and three-quark contributions: \\
\begin{equation}
  \widehat{r}_B^{2} = A\sum_{i}e_i.1 + B\sum_{i\neq j} e_i \sigma_{i}\sigma_{j} + C\sum_{i\neq j\neq k}e_i \sigma_{j}\sigma_{k}
\end{equation}
For octet and decuplet baryons, the above expression of charge radii can be written as:
\begin{equation}
\widehat{{r}}^{2}_{1/2} = (A-3B)\sum_{i}e_i + 3(B-C)\sum_{i} e_i\sigma_{iz}   
\end{equation}
\begin{equation}
   \widehat{{r}}^{2}_{3/2} = (A-3B+6C)\sum_{i}e_i + 5(B-C)\sum_{i} e_i\sigma_{iz}  
\end{equation}
Here $\widehat{{r}}^{2}_{1/2}$ and $\widehat{{r}}^{2}_{3/2}$represents the charge radii operator for spin- $\frac{1}{2}$ and spin-$\frac{3}{2}$ particles respectively. The unknown parameters A, B and C are introduced in the operator's expansion, parametrizing the contribution of orbital and color space \cite{29,30}.\\
Using the operator formalism, the quadrupole moment and charge radii of J$^P$=$\frac{1}{2}^+$ octet and J$^P$=$\frac{3}{2}^+$ decuplet baryons is calculated. The matrix element of the operator corresponding to the three-quark spin-flavor wavefunction is evaluated. For example: the charge radii of $\Sigma^{*+}$ expressed as:\\
\begin{equation}
    \begin{aligned}
       \langle\Phi_{3/2}^{(\uparrow)}|{\widehat{O}}| \Phi_{3/2}^{(\uparrow)}\rangle =& \frac{1}{N^2} [{a_0}^{2}{\langle \Phi_{1}}^{(\frac{3}{2} \uparrow)}|\widehat{O}|{\Phi_{1}}^{(\frac{3}{2}\uparrow)}\rangle + {b_1}^2 \langle{\Phi_1}^{(\frac{3}{2}\uparrow)}|\widehat{O}|{\Phi_1}^{(\frac{3}{2}\uparrow)}\rangle + \\& {b_8^2 \langle{\Phi_{8}}^{(\frac{1}{2}\uparrow)}|\widehat{O}|{\Phi_{8}}^{(\frac{1}{2}\uparrow)}\rangle + {d_1}^{2}\langle{\Phi_1}^{(\frac{3}{2}\uparrow)}|\widehat{O}|{\Phi_1}^{(\frac{3}{2}\uparrow)}\rangle + {d_8}^2 \langle{\Phi_8}^{(\frac{1}{2}\uparrow)}|\widehat{O}|{\Phi_8}^{(\frac{1}{2}\uparrow)}\rangle}    
    \end{aligned}
\end{equation}
Here, $\widehat{O}$ represents the charge radii operator of decuplet baryon mentioned in eq. (17). After applying the operator, we get the expression in terms of the statistical coefficients $( a_0, a_8, a_{10}, b_1, b_8, b_{10}, c_8, d_8 )$ and the parameters A, B $\&$ C.\\
\begin{equation}
\begin{aligned}
       &\frac{1}{N^2}[ a_0^2 (0.999 A + 1.998 B + 0.999 C)+ b_1^2(- 0.9637 A + 7.0022 B -9.8934 C )+\\& {b_8^2} (0.2101 A + 4.036 B - 3.4060 C) + d_1^2 (0.2 A + 0.4 B + 0.2 C) + d8^2 ( 0.5333 A - 1.1 B + 2.7 C)] 
\end{aligned}
\end{equation}
Similar expressions are obtained for all octet and decuplet baryons in SU(3) symmetry limit. The statistical coefficients are associated with the probability of each state in flavor, spin and color space individually. The statistical approach in conjugation with the principle of detailed balance used to determine the set of probabilities in every aspect 
\begin{table}[ht!]
     \centering
     \small\addtolength{\tabcolsep}{0.5pt}
     \begin{tabular}{p{2 cm}p{5.5 in}}\hline
        Charge radii & \hspace{6cm}Related expressions\\\hline
       \hspace{2mm} $r^{2}_{\Sigma^{*+}}$&$a_{0}^2$ (0.7216 - 0.1804 r) + $b_{1}^2$ (3.8221 - 0.7780 r) + $b_{8}^2$ (1.7223 - 0.1643 r) + $d_{1}^2$ (0.1444 - 0.0361 r) - $d_{8}^2$ (0.7684- 0.1921 r)\\
    \hspace{3mm}$r^{2}_{\Sigma^{*-}}$&  $a_{0}^2$ (-0.3612 - 0.1806 r) + $b_{1}^2$ (-1.9111 - 0.7780 r) + $b_{8}^2$ (-0.8611 - 0.1643 r) + $d_{1}^2$ (-0.0722 - 0.0361 r) + $d_{8}^2$ (0.3842 + 0.1921 r)\\
       \hspace{3mm}$r^{2}_{\Sigma^{*0}}$& $a_{0}^2$ (0.1806 - 0.1806 r) + $b_{1}^2$ (0.2939 - 0.4714 r) + $b_{8}^2$ (0.4790 - 0.4790 r) + $d_{1}^2$ (0.0361 - 0.0361 r) - $d_{8}^2$ (0.2094 - 0.2094 r)\\\hline
        \hspace{3mm}$r^{2}_{\Xi^{*-}}$ & $a_{0}^2$ (-0.1806 - 0.3612 r) + $b_{1}^2$ (-0.2293 - 0.8136 r) + $b_{8}^2$ (-0.1643 - 0.8611 r) + $d_{1}^2$ (-0.1083 - 0.2167 r) + $d_{8}^2$ (0.1921 + 0.3842 r)\\ 
       \hspace{2mm} $r^{2}_{\Xi^{*0}}$ & $a_{0}^2$ (0.3612 - 0.3612 r)  +$b_{1}^2$ (1.685 - 2.0404 r) + $b_{8}^2$ (0.4256 - 0.9581 r) + $d_{1}^2$ (0.0722 - 0.0722 r) - $d_{8}^2$ (0.3842 - 0.3842 r)\\ \hline
       \hspace{3mm} $r^{2}_{\Omega^{-}}$& $a_{0}^2$ (-0.5418 r) + $b_{1}^2$ (- 1.2204 r) + $b_{8}^2$ (-0.7128 - 0.2007 r) + $d_{1}^2$ (0.1503 r) - $d_{8}^2$ (-0.0846 + 0.6124 r) \\\hline
       \hspace{2mm} $r^{2}_{\Sigma^{+}}$& ${a_0}^2$ (0.7851 + 0.0699 r) + ${a_8}^2$ (1.1369 + 0.2125 r) + ${a_{10}}^2$ (0.2527 - 0.0631 r) + ${b_1}^2$ (0.3136 - 0.1671 r) +  ${b_8}^2$ (0.1598 - -0.2055 r) + ${b_{10}}^2$ (0.4911 - 0.1227 r) + ${c_8}^2$ (1.1691 - 0.2479 r) + ${d_8}^2$ (-0.4819 + 0.0938 r)\\
        \hspace{3mm}$r^{2}_{\Sigma^{-}}$& ${a_0}^2$ (-0.3925 + 0.0699 r) + ${a_8}^2$ (-0.5684 + 0.2125 r) + ${a_{10}}^2$ (-0.1263 - 0.0631 r) + ${b_1}^2$ (-0.1568 - 0.1671 r) +  ${b_8}^2$ (-0.0799 -0.2055 r) + ${b_{10}}^2$ (-0.2455 - 0.1227 r) + ${c_8}^2$ (-0.5845 - 0.2479 r) + ${d_8}^2$ (0.2409 + 0.0938 r)\\ 
       \hspace{3mm}$r^{2}_{\Sigma^{0}}$& ${a_0}^2$ (0.0534 - 0.1466 r) + ${a_8}^2$(0.4411 -0.5486 r) +${a_{10}}^2$ (0.06319 - 0.1963 r) + ${b_1}^2$ (0.10757 -  0.08199 r) +  ${b_8}^2$ (0.28425 - 0.30705 r) + ${b_{10}}^2$(0.09298 - 0.04861 r) + ${c_8}^2$ (0.31448  - 0.2738 r) + ${d_8}^2$ (-0.0627 + 0.1093 r)\\\hline
        \hspace{3mm}$r^{2}_{\Xi^{-}}$ & ${a_0}^2$ (0.0699 - 0.3925 r) + ${a_8}^2$ (0.2125 - 0.5684 r) + ${a_{10}}^2$ (-0.0631 - 0.1263 r) + ${b_1}^2$ (-0.1671 - 0.1568 r) +  ${b_8}^2$ (-0.2055 - 0.0799 r) + ${b_{10}}^2$ (-0.1227 - 0.2455 r) + ${c_8}^2$ (-0.2479 - 0.5845 r) + ${d_8}^2$ (0.0938 + 0.2030 r)\\ 
       \hspace{2mm} $r^{2}_{\Xi^{0}}$ & ${a_0}^2$ (-0.1398 - 0.3925 r) + ${a_8}^2$ (-0.4251 - 0.5684 r) + ${a_{10}}^2$ (0.1263 - 0.1263 r) + ${b_1}^2$ (0.3343 - 0.1568 r) +  ${b_8}^2$ (0.4111 - 0.0799 r) + ${b_{10}}^2$ (0.2455 - 0.2455 r) + ${c_8}^2$ (0.4958 - 0.5845 r) + ${d_8}^2$ (-0.1877 + 0.2030 r)\\ \hline
       \hspace{3mm} $r^{2}_{\Lambda^{0}}$&
  ${a_0}^2$ (0.05340 - 0.1466 r) + ${a_8}^2$(0.4411 - 0.54868 r) + ${a_{10}}^2$ (0.06319 - 0.19631 r) +${b_1}^2$(0.10757 - 0.08199 r) + ${b_8}^2$ (0.2842 - 0.30705 r) + ${b_{10}}^2$ (0.09298 - 0.0486 r) + ${c_8}^2$ (0.3144 - 0.2738 r) + ${d_8}^2$ (-0.0627 + 0.1093 r) \\\hline
        \end{tabular}
        \caption{Expression obtained after applying charge radii operator to the baryon octet $J^P = \frac{1}{2}^+$ and decuplet $J^P = \frac{3}{2}^+$ particles}
     \label{tab:1}
 \end{table}\\
i.e. flavor, spin and color space which is discussed in the next section. The parameters A, B, and C are calculated previously in our article Ref. \cite{53} using available experimental data for charge radii and quadrupole moment. It should be noted that the flavor SU(3) symmetry is an approximate symmetry, since u, d, and s quarks have different masses which breaks SU(3) symmetry. To investigate the symmetry breaking effect in valence, a mass dependent parameter 'r' is introduced in the operator. After applying the operator, the eigenvalues obtained have the form of seven statistical coefficients and symmetry breaking operator 'r'. 
The obtained expressions for charge radii and quadrupole moment are tabulated in Tables 1 and 2 respectively.
 \begin{table}[ht!]
     \centering
     \small\addtolength{\tabcolsep}{0.5pt}
     \begin{tabular}{p{2 cm}p{5.5 in}}\hline
        Quadrupole moment & \hspace{6cm}Related expressions\\\hline
       \hspace{2mm} $Q_{\Sigma^{*+}}$ & $a_{0}^2$ (-0.2586 + 
       0.06466 r) + $b_{1}^2$ (- 0.4980 + 0.1096 r) + $b_{8}^2$ (-0.48117 + 0.09802 r) + $d_{1}^2$ (0.0485 + 0.0364 r) - $d_{8}^2$ (0.3898 - 0.04893 r)\\
      \hspace{3mm}$Q_{\Sigma^{*-}}$ & $a_{0}^2$ (0.1293 + 0.0646 r) + $b_{1}^2$ (0.2490 + 0.1096 r) + $b_{8}^2$ (0.2405 + 0.0980 r) + $d_{1}^2$ (-0.0242 + 0.0364 r) - $d_{8}^2$ (-0.1949 - 0.074 r)\\ 
       \hspace{3mm}$Q_{\Sigma^{*0}}$ & $a_{0}^2$ (-0.0646 + 0.0646 r) + $b_{1}^2$ (0.01582 - 0.0549 r) + $b_{8}^2$ (-0.0018 + 0.0358 r) + $d_{1}^2$ (0.0137 - 0.05284 r) - $d_{8}^2$ (-0.0212 + 0.02941 r)\\\hline
        \hspace{3mm}$Q_{\Xi^{*-}}$ &  $a_{0}^2$ (0.06466 + 0.1293 r) + $b_{1}^2$ (0.1096 + 0.24902 r) + $b_{8}^2$ (0.0980 + 0.24059 r) + $d_{1}^2$ (0.0364 - 0.0242 r) - $d_{8}^2$ (0.074 - 0.19493 r)\\ 
       \hspace{2mm} $Q_{\Xi^{*0}}$ & $a_{0}^2$ (-0.1293 + 0.1293 r) + $b_{1}^2$ (-0.2193 + 0.2490 r) + $b_{8}^2$ (-0.1960 + 0.24059 r) + $d_{1}^2$ (-0.1762 + 0.0792 r) - $d_{8}^2$ (0.0978 - 0.19493 r)\\\hline
       \hspace{3mm} $Q_{\Omega^{-}}$& $a_{0}^2$ (0.194 r) + $b_{1}^2$ (0.37113 r) + $b_{8}^2$ (0.35728 r) + $d_{1}^2$ (-0.13826 r) - $d_{8}^2$ (-0.2976 r)\\\hline
        \hspace{3mm}$Q_{\Sigma^{+}}$& ${a_0}^2$ (0.00206 r) + ${a_8}^2$ (0.0107 + 0.0011 r) + ${a_{10}}^2$ (-0.00827 - 0.00206 r) + ${b_1}^2$ (0.1489 + 0.04320 r) +  ${b_8}^2$ (0.22007 + 0.06114 r) + ${b_{10}}^2$ (- 0.01608 + 0.00402 r) + ${c_8}^2$ (-0.3434 + 0.10647 r) + ${d_8}^2$ (-0.0223 + 0.0418 r)\\ 
        \hspace{2mm} $Q_{\Sigma^{-}}$& ${a_0}^2$ (0.00206 r) + ${a_8}^2$ (-0.00537 + 0.0011 r) + ${a_{10}}^2$ (0.00413 + 0.00206 r) + ${b_1}^2$ (-0.07446 + 0.04320 r) +  ${b_8}^2$ (-0.14651 + 0.07684 r) + ${b_{10}}^2$ (0.00804 + 0.00402 r) + ${c_8}^2$ (0.1717 + 0.10647 r) + ${d_8}^2$ (0.0111 + 0.04188 r)\\
       \hspace{3mm}$Q_{\Sigma^{0}}$& ${a_0}^2$ (-0.01904 + 0.0240 r) + ${a_8}^2$ (-0.03312 + 0.03491 r) + ${a_{10}}^2$ (-0.03249 + 0.03145 r) + ${b_1}^2$ (- 0.02099  + 0.00311 r) + ${b_8}^2$ (- 0.03966 + 0.00666 r) + ${c_8}^2$ (0.09844 + 0.10230 r) + ${d_8}^2$ (-0.09654 + 0.10009 r)\\\hline
        \hspace{3mm}$Q_{\Xi^{-}}$ & ${a_0}^2$ (0.00206 r) + ${a_8}^2$ (0.00117 - 0.0053 r) + ${a_{10}}^2$ (0.0020 + 0.00413 r) + ${b_1}^2$ (0.0432 - 0.07446 r) +  ${b_8}^2$ (0.07885 - 0.14249 r) + ${b_{10}}^2$ (0.00402 + 0.00804 r) + ${c_8}^2$ (0.1064 + 0.1717 r) + ${d_8}^2$ (0.04188 + 0.0111 r)\\ 
       \hspace{2mm} $Q_{\Xi^{0}}$ & ${a_0}^2$ (-0.00413) + ${a_8}^2$ (-0.0023 - 0.00537 r) + ${a_{10}}^2$ (-0.0041 + 0.00413 r) + ${b_1}^2$ (-0.0864 - 0.0744 r) +  ${b_8}^2$ (-0.1517 - 0.14844 r) + ${b_{10}}^2$ (-0.0080 + 0.0080 r) + ${c_8}^2$ (-0.2129 + 0.1717 r) + ${d_8}^2$ (-0.0837 + 0.0111 r)\\ \hline
       \hspace{3mm} $Q_{\Lambda^{0}}$& ${a_0}^2$ (-0.01904 + 0.02402 r) + ${a_8}^2$ (-0.03312 + 0.03491 r) + ${a_{10}}^2$ (-0.03249 + 0.03145 r) + ${b_1}^2$ (-0.02099  + 0.00311 r) + ${b_8}^2$ (- 0.03966 + 0.00666 r) + ${c_8}^2$ (0.09844 + 0.10230 r) + ${d_8}^2$ (-0.09654 + 0.10009 r)  \\\hline
        \end{tabular}
        \caption{Expression obtained after applying quadrupole moment operator to the baryon octet $J^P = \frac{1}{2}^+$ and decuplet $J^P = \frac{3}{2}^+$ particles}
     \label{tab:2}
 \end{table}
\section {Principle of detailed balance and Statistical model}
The principle of detailed balance is proposed by Zhang et al. \cite{54}, based on the assumption that each physical hadron state can be expanded as an ensemble of quark-gluon Fock states. Each Fock states avail the presence of quark-antiquark pairs multi-connected through gluons and it can be expressed as: \\
\begin{equation}
|h \rangle = \sum_{i,j,k,l} C_{i,j,k,l} | \{q^3\},\{i,j,k,l\}\rangle
\end{equation}
Here $\{q^3\}$ represents the constituent quarks of baryons, i, j and l is the number of quark-antiquark pairs i.e. $u\bar{u}$, $d\bar{d}$, $s \bar s$ pairs respectively and k is the number of gluons. Basically, the quarks and gluons present in the Fock states are the intrinsic partons which are multiconnected non-perturbatively to the valence quarks \cite{55}. The probability of finding the baryon in quark-gluon Fock state $|\{q^3)\},\{i,j,k,l\}\rangle$ expressed as:
\begin{equation}
  \rho_{i,j,k,l} = |C_{i,j,k,l}|^2 
\end{equation} \\where $\rho_{i,j,k,l}$ satisfy the condition of normalization $\sum_{i,j,k,l} \rho_{i,j,k,l} = 1$.\\
This principle considered that any two nearby quark-gluon Fock states should balance with each other \cite{56}. This means the probability of finding the baryon in any Fock state should not change during the time which is expressed as: \\
$$\rho_{i,j,k,l} |\{q^{3}\},\{i,j,k,l\}\rangle \rightleftharpoons \rho_{i',j',k',l'}|\{q^{3}\},\{i',j',k',l'\}\rangle$$\\
The transition probability of different Fock state in flavor space is calculated with the help of various sub-processes like g $\rightleftharpoons  q \bar q$ , g $\rightleftharpoons $ gg, and q $\rightleftharpoons $ qg. By taking the strange quark into consideration, the whole scenario is modified due to its large mass. 
To have transition processes like g $\Leftrightarrow s \bar s$, the gluons must possess sufficient free energy greater than at least two times the mass of strange quark i.e. $\varepsilon_g > 2M_s$, where $M_s$ is the mass of strange quark. The emergence of $s \bar s$ pairs from gluons is restricted by a constraint appears in the form of $k(1 - C_{l})^{n-1}$ \cite{57}, where n is the total number of partons present in the Fock state, $l$ is the no. of $s \bar s$ pairs and k is the no. of gluons. The constraints arise from the gluon free energy distribution and total energy of partons present in the baryon. For all cases, the value of $C_{l-1}=\frac{2M_s}{M_B-2(l-1)M_s}$, where $M_B$ is the mass of baryon. It is important to note that the value of constraint $(1 - C_{l})^{n-1}$ clearly indicates the difference between double strange baryon and single strange baryon to accommodate $s \bar s$ pairs \cite{58}. With the increasing no. $\bar s s$ pairs in sea with doubly strange baryons, the value of $(1 - C_{l})^{n-1}$ decreases and it affects the overall probability of Fock states. The Fock states without strange quark content cover 86$\%$ of the Total Fock states while the inclusion of $s \bar s$ reduce it to 80$\%$ \cite{59}. The interesting part of taking the strange quark is that the splitting and recombination for the processes g $\Leftrightarrow s \bar s$ experience a breaking in SU(3) symmetry within the sea.  
The detailed calculation of probabilities including the strange quark content in sea given in Ref. \cite{55,58,60}.\\ Now, the probability in flavor space is further needed to calculate probabilities in spin and color space. The statistical decomposition of baryonic state in various Fock states like $|u \bar ug\rangle$, $|d \bar dg\rangle$, $|s \bar sg\rangle$, $|u \bar u d\bar d\rangle$, $|u \bar u d\bar dg\rangle$ is used to find probabilities in spin and color space with their appropriate multiplicity. The multiplicities for all Fock states are calculated in the form of $\rho_{p,q}$ where the relative probability corresponds to the ’Spin p’ for the valence part and ’Spin q’ for sea components. This ensures that the resultant spin should come out as 1/2 for octet baryons and 3/2 for decuplet baryons. Similarly, the probabilities for the color spaces can be calculated which yields the color singlet state. The detailed calculations of these multiplicities are discussed in ref. \cite{55,58,59}. The sum of the total probabilities in spin, flavor and color space will give the coefficients $a_0, a_8, a_{10}, b_1, b_8, b_{10}, c_8, d_8$ to the total wavefunction. The scalar, vector and tensor sea contribution are expressed in terms of these coefficients. The statistical coefficients offer insights into how 'sea quarks' contribute to diverse properties of baryons, including masses \cite{61}, semi-leptonic decays \cite{62}, charge radii\cite{53} etc. The combination of the principle of detailed balance and statistical approach emerges as a reliable method for characterizing the various properties 
of octet and decuplet baryons.
\section {Numerical results and discussion}
Probing the internal configuration of the ground and excited state baryons, the charge radii and electric quadrupole moment are the most interesting observables. To compute these properties, a suitable operator (discussed in section 3) is applied to the baryonic wavefunction. Our aim is to explore the significance of sea in the relative probabilities of the Fock states containing both strange and non-strange quark contents. The statistical coefficients ($a_0, a_8, a_{10}, b_1, b_8, b_{10}, c_8, d_8$) represent the individual contributions from various parts of sea categorized as scalar (spin-0), vector (spin-1) and tensor (spin-2) depending upon their spin. In order to examine the contribution of scalar sea alone, we suppress the contribution of vector and tensor sea. The value of coefficients assumed b$_{1,8,{10}}$, c$_8$, d$_8$ =0 for scalar sea, the coefficients a$_{0,8,{10}}$, d$_8$=0  for vector sea and for tensor sea the coefficients taken as a$_{0,8,{10}}$, b$_{1,8,{10}}$, c$_8$ =0 for octet baryons. A similar analysis applied to check the contribution of individual sea for decuplet baryons. 
Due to the large mass difference of strange (s) and non-strange quarks (u,d), we also studied the SU(3) symmetry and breaking effect on charge radii and quadrupole moments. To analyze the SU(3) breaking effect in valence, the relevant operator directly involves the strange quark mass in the form of the parameter ’r’. The parameter 'r' is defined as: r = $\frac{\mu_s}{\mu_d}$ \cite{35} where $\mu_s$ is the magnetic moment of the strange quark and $\mu_s$ is the magnetic moment of the d quark. A direct dependence of $m_s$ and $m_{u/d}$ with some constant factors involved in the parameter 'r'. Strange sea also becomes an active participant with the direct inclusion of the strange mass correction. This correction modifies the probabilities of Fock states, leading to changes in the statistical coefficients. The statistical approaches include various models i.e. Model C, P, and D that enable us to explore the influence of sea dynamics to the several static properties of baryons \cite{55}. Model C is the basic model that contains various quark-gluon Fock states and assumes the equal probability of each Fock state. Model D is the modified picture of Model C. It involves the suppression of Fock states associated with higher multiplicities introduced by Sing and Upadhyay \cite{55}. Sea with greater multiplicity in color and spin space has less probability of survival due to higher interactions. We compute the charge radii and quadrupole moment using Model D, presented in Tables 3 and 4. It is important to note that the best-fit value of 'r' is obtained by using a suitable fitting algorithm and found to be r = 0.850. Also, the parameters mentioned in operators i.e. A, B, and C are used as input which is calculated previously in Ref. \cite{53} by $\chi^2$ minimization method.  
\subsection{Charge radii}
The charge radii hold interest in providing vital information on the spatial distribution of charge inside the baryons. It is influenced by various factors such as masses of the constituent quarks and their specific flavor composition. 
Using the statistical framework, the numerical values of electric charge radii for octet $J^P = \frac{1}{2}^+$ and decuplet $J^P = \frac{3}{2}^+$ particles are presented in Table 3. 
The statistical parameters ($a_0, a_8, a_{10}, b_1, b_8, b_{10}, c_8, d_8$) provide the significant contribution of the strange and non-strange $q\bar q$ condensates.\\
$\mathbf{(a)}$ $\mathbf{J^P= \frac{3}{2}^+ baryons}$\\
For decuplet particles ( $\Sigma^{*-}$, $\Sigma^{*+}$, $\Xi^{*-}$, $\Omega^{*-}$), the scalar sea acts as a major contributor from the total sea. One reason for this is possibly the larger multiplicities of valence quark spin states when combined with the sea quarks spin (i.e. spin - 0, 1, and 2). The probability of having spin- $\frac{3}{2}$ is more when coupled with spin-0 (scalar sea) as compared to spin - 1,2 (vector, tensor sea). Although, the vector and tensor sea contribution is less but it cannot be neglected. In SU(3) symmetry limit, the contribution of pure scalar sea to the charge radii is about more than $90\%$ whereas the vector and tensor sea contributed nearly $ 2-8\%$. In the case of neutral particles i.e. $\Xi^{*0}$,  $\Sigma^{*0}$ only vector sea contribute while the scalar and tensor sea impact is zero. This is because the contribution of valence quark wavefunction across the scalar sea ($a_0$) and tensor sea ($d_1, d_8$) terms are zero. 
If strange sea is assumed then following points are observed to the forefront:\\
1. For $\Sigma^{*-}$(dds), $\Xi^{*-}$(dss), $\Omega^{*-}$(sss) baryons, a decrease in charge radii value is observed in Table 3. This may be possible due to the larger mass of strange quark that limit the free energy of gluon, which affects the emission of virtual gluon. As a result, the Fock states with large no. of $s\bar s$ pairs in sea are assumed to be less probable. 
It also indicates that charge radii of baryons are directly influenced by the probabilities associated with accommodating $s\bar s$ pairs. 
On the other side, the value of charge radii is increased for $\Sigma^{*+}$(uus), $\Sigma^{*0}$(uds), $\Xi^{*0}$(uss) baryons. In addition to the quark-mass dependence, the electric charge of quarks is also a crucial factor that influences the charge radii of baryons. Due to the high charge of up(u) quark, it may be possible that the strength of electromagnetic interaction dominates for $u\bar u$ as compared to $d\bar d, s\bar s$ condensates. 
\\
2. For neutral particles i.e. $\Sigma^{*0}, \Xi^{*0}$, the scalar sea becomes an active contributor, as the effect of vector and tensor sea is much less and it can be neglected. Due to the non-negligible mass of strange quark, a mass correction parameter 'r' is considered for valence. Because of this, the valence quark contribution across the scalar sea ($a_0$) cannot be zero. In SU(3) symmetry limit, the decuplet baryon charge radii can be expressed with the following relations:
$$r^{2}_\Sigma{*+}= r^{2}_\Sigma{*-}=r^{2}_\Xi{*-}= r^{2}_\Omega{*-}$$
While the inclusion of symmetry breaking changes the pattern considerably and we get
$$r^{2}_\Sigma{*+} > r^{2}_\Sigma{*-}>r^{2}_\Xi{*-}>r^{2}_\Omega{*-}$$
The charge radii value deviates $4-10\%$ when compared with SU(3) symmetry results. We observed a considerable change in both magnitude and sign of charge radii of neutral strange baryons ($\Sigma^{*0}, \Xi^{*0}$) after incorporating the strange sea. The change in the charge radii of $\Sigma^{*0}, \Xi^{*0}$ are closely matched with Lattice QCD predictions \cite{25,26}.\\
$\mathbf{(b)}$ $\mathbf {J^P=\frac{1}{2}^+ baryons}$\\
In case of octet particles, the sea is found to be dynamic for scalar plus vector sea within SU(3) symmetry. The tensor sea contribution is negligible for all the baryons because of the quark spin-flip processes. The contribution of the scalar sea is about $50-80\%$ whereas the vector sea contributed about $ 15-50\%$ to the charge radii of $J^P=\frac{1}{2}^+$ particles. 
The neutral particles ($\Sigma^0, \Lambda^0$) has zero valence quark contribution. Further, when the strange sea is considered, the contribution of individual sea (scalar, vector, and tensor) exhibits similar dominancy with the SU(3) symmetry. The interesting point is observed that the charge radii value is decreased for doubly strange particles ($\Xi^0, \Xi^-$) and increased for singly strange baryons ($\Sigma^+, \Sigma^-$). 
It may be possible because when doubly strange baryons accommodate large no. of $s\bar s$ pairs in the sea, the value of suppression factor $(1-C_l)^{n-1}$ decreases, and it automatically decreases the probability of that particular Fock state as discussed in the preceding section. The chances of the interaction of sea having $s \bar s$ pairs with heavier strange baryons are less as compared to lighter strange baryons. Moreover, within SU(3) symmetry limit we get the following pattern for the octet charge radii :
$$r^{2}_{\Sigma^{+}} > r^{2}_{\Xi^{-}}>r^{2}_{\Xi^{0}}>r^{2}_{\Sigma^{-}}$$ 
But the results are affected by the inclusion of SU(3) symmetry breaking and give:
$$r^{2}_{\Sigma^{+}} > r^{2}_{\Sigma^{-}}>r^{2}_{\Xi^{-}}>r^{2}_{\Xi^{0}}$$  
The importance of the sea lies in the suppression of higher multiplicities of the quark–gluon
Fock states. However, in the case when the sea is excluded in the statistical model, the charge
radii of both octet and decuplet baryons deviate more than 50$\%$ from the computed value. In comparison, our predicted value of charge radii shows good agreement in sign and magnitude with different phenomenological models \cite{25,26,27,28,30}.
\subsection{Quadrupole moment}
The quadrupole moment is an important structural property that characterizes the charge distribution within the baryons and gives insights into the deformities in the shape of the baryon. This deformation is related to the angular momentum and spin of the valence quarks. The statistical model is sensitive to the probabilities associated with spin, flavor, and color space for both valence and sea quarks. By using this approach, we computed the numerical results of the quadrupole moment of J$^P=\frac{1}{2}^+,\frac{3}{2}^+$ baryons which is presented in Table 4.\\
$\mathbf{(a)}$ $\mathbf{J^P= \frac{3}{2}^+ baryons}$\\  For decuplet particles ( $\Sigma^{*-}$, $\Sigma^{*+}$, $\Xi^{*-}$, $\Omega^{*-}$), the scalar sea acts as a major contributor from the total sea. 
For neutral particles i.e. $\Xi^{*0}$,  $\Sigma^{*0}$ only tensor sea contributed while the scalar and vector sea impact is negligible. The vector and tensor contributed 4-10$\%$ whereas the scalar sea contribution is about $88\%$ to the quadrupole moment.
This might indicate that the chances of having spin- $(\frac{3}{2})^{+}$ is more with the scalar sea (spin-0) as compared to vector and tensor sea. On corporating the effect of strange sea, a few points need to be addressed.\\ 
1. The value of the quadrupole moment is decreased for negatively charged particles i.e. $\Sigma^{*-}$(dds), $\Xi^{*-}$(dss), $\Omega^{*-}$(sss) 
and increased for $\Sigma^{*+}$(uus), $\Sigma^{*0}$(uds), $\Xi^{*0}$(uss) baryons. The variation in the value of quadrupole moment is influenced by the splitting and recombination between gluons and $s\bar s$ pairs undergoing the process $g \Leftrightarrow s\bar s$. Instead of this, the breaking parameter 'r', associated with the strange baryons, contributes to the change in the quadrupole moment value. From the results, we can suggest that the charge distribution inside the baryons is symmetric or in a compact manner for the negatively charged particles ($\Sigma^{*-}$, $\Xi^{*-}$, $\Omega^{*-}$) as compared to other baryons.\\
2. For neutral ($\Sigma^{*0}, \Xi^{*0}$) particles, the dominancy of pure scalar can be easily observed from Table 4. The impact of vector and tensor sea can be neglected. In SU(3) symmetry limit, the magnitude of quadrupole moments of decuplet baryons can be expressed by the following relations:
$$Q_{\Sigma^{*+}}= Q_{\Sigma^{*-}}=Q_{\Xi^{*-}}= Q_{\Omega^{*-}}$$
The symmetry breaking effect changes the pattern considerably and we get
$$ Q_{\Sigma^{*-}}> Q_{\Xi^{*-}}> Q_{\Omega^{*-}}>Q_{\Sigma^{*0}}>Q_{\Xi^{*0}}>Q_{\Sigma^{*+}}$$
In statistical model, we predicted an oblate shape for the $\Sigma^{*+}, \Sigma^{*0}, \Xi^{*0}$ baryons and a prolate shape for $\Sigma^{*-}, \Xi^{*-}, \Omega^{*-}$. On comparison with SU(3) symmetry results, the quadrupole moment value deviates upto 14$\%$ for $\Sigma^{*-},\Sigma^{*+},\Xi^{*-},\Omega^{*-}$ baryons and the deviation is maximum for neutral baryons ($\Sigma^{*0}, \Xi^{*0}$) which is more than 80$\%$. This signifies the impact of strange sea quarks with strange baryons. \\
$\mathbf{(a)}$ $\mathbf{J^P= \frac{1}{2}^+ baryons}$\\ 
The emission of virtual gluons dominates the sea, suggesting that the vector sea coefficients $b_1, b_8, c_8$ are likely to be more dominating. From Table 4, we can observe the vector sea dominancy to the quadrupole moment value. The contribution of vector sea is more than $95\%$. For each octet particle, scalar and tensor sea contribution is is nearly negligible. Neglecting the tensor sea is based on the fact that tensor sea contribution comes from spin- $ \frac{3}{2}$ valence part and the probability for the core part to have spin- $\frac{3}{2}$ is very less. Neutral baryons, i.e. $\Sigma^0, \Lambda^0$, exhibit zero value of quadrupole moment. The magnitude of quadrupole moment within the SU(3) symmetry limit can be expressed as:
$$Q_{\Sigma^{+}} > Q_{\Xi^{0}}>Q_{\Xi{-}}>Q_{\Sigma^{-}}$$
Also, we have
$$Q_{\Sigma^{0}}=Q_{\Lambda^{0}}$$
where the three-quark core does not contribute to the quadrupole moment. On accounting for the strange sea, the contribution of individual sea (scalar, vector and tensor) exhibits similarity with the non-strange sea. Due to symmetry breaking effect, we observed that the quadrupole moment decreases for $\Xi^-$(dss) and increases for $\Sigma^-$(dds). This is because higher masses of quarks have lesser chances than the lighter quarks due to the limited energy of gluon. On the contrary, we note an increase in the quadrupole moment for $\Xi^0$(uss) and a decrease for $\Sigma^+$(uus). 
The magnitude of the quadrupole moment with symmetry breaking effect can be expressed by the following relations:
$$Q_{\Sigma^{+}} > Q_{\Xi^{0}}>Q_{\Xi^{-}}>Q_{\Sigma^{0}}>Q_{\Sigma^{-}}>Q_{\Lambda^{0}}$$
A non-vanishing value for the neutral baryons ($\Sigma^0, \Lambda^0$) is generated by the mass correction parameter 'r' and sea quarks. The quadrupole moment value deviates upto $30\%$  with the maximum deviation exceeding 50$\%$ for $\Sigma^{-}$. 
When the sea is completely excluded in the statistical model, the quadrupole moment for both $J^P= \frac{1}{2}^+$ and $J^P= \frac{3}{2}^+$ baryons deviates more than 50$\%$ from the computed value. Further, using the statistical approach we predicted a prolate shape for $\Sigma^{+}, \Sigma^{-}, \Xi^{-}$ baryons while an oblate shape for $\Sigma^{0}, \Xi^{0}, \Lambda^{0}$ neutral baryons.  Our computed results are consistent with various phenomenological models \cite{29,30,27}. Since there is no experimental information available on the quadrupole moment, the accuracy of our statistical results can be assessed by future experiments.
\section {Conclusion}
In our present work, the statistical approach is used to calculate the electromagnetic properties i.e. charge radii and quadrupole moment of strange octet and decuplet baryons. Baryons are assumed to have a virtual dynamic sea with strange and non-strange $q\bar q$ condensates in addition to gluons. Our main attention is to find the impact of strange and non-strange partons on the above-mentioned properties. Since the sea contains quark-gluon Fock states, the principle of detailed balance is used to determine the probabilities for these Fock states in terms of statistical parameters ($a_{0}, a_{8}, a_{10}, b_{1}, b_{8}, b_{10}, c_{8}, d_{8}$). To appreciate the strange sea, a strangeness suppression factor $(1 - C_{l})^{n-1}$ is discussed which modifies all the probabilities related to different Fock states. It also induces a breaking in SU(3) symmetry within the sea.  
To study the breaking effect in valence, the relevant operator directly involved the strange quark mass in the form of the parameter 'r'. Our analysis concluded that, for $J^P= \frac{3}{2}^+$ particles, the scalar sea is the major contributor to both properties. In the case of $J^P= \frac{1}{2}^+$ particles, the charge radii are mainly influenced by the scalar plus vector sea, while the quadrupole moment is primarily dominated by the vector sea. The strange sea seems to contribute effectively, there by suggesting the gluons in the sea to be generating strange quark-antiquark condensates. 
The statistical approach predicts an oblate shape for $\Sigma^{*+}, \Sigma^{*0}, \Xi^{*0}$, $\Sigma^{0}, \Xi^{0}, \Lambda^{0}$ baryons and a prolate shape for $\Sigma^{+}, \Sigma^{-}, \Xi^{-}$, $\Sigma^{*-}, \Xi^{*-}, \Omega^{*-}$ baryons. Also, the analysis of the quadrupole moment concludes that the strange sea quarks have a pronounced impact on the overall structure of baryons without directly inducing any changes in their shape. 
We conclude that the strange sea provides a better SU(3) analysis for charge radii and quadrupole moment. Its contribution plays a significant role in determining the validity of the present approach. It is worth mentioning that our calculations are performed for the scale of order 1GeV$^2$ and are in a non-relativistic
frame.
\section{Acknowlegement}
The authors gratefully acknowledge the financial support by the Department of Science and Technology (SERB/F/9119/2020), New Delhi.
\begin{landscape}
\begin{table*}[h!]{\normalsize
\renewcommand{\arraystretch}{1.0}
\tabcolsep 0.1cm
    \centering
   \small{
   \begin{tabular}{cccccc|cccccccccc}\toprule\hline
           &           & \multicolumn{8}{c}{Statistical model} &  &\\\cmidrule{3-10}
   Baryons & \Longunderstack{Charge\\ radii}&\Longunderstack{SU(3) Symmetry \\with \\(scalar+vector\\+tensor) sea}& \Longunderstack{Scalar \\Sea}& \Longunderstack{Vector\\ Sea}& \Longunderstack{Tensor\\ Sea} & 
   \Longunderstack{SU(3) symmetry\\ breaking with\\(Valence+sea)}  & \Longunderstack{Scalar\\Sea}& \Longunderstack{Vector\\Sea}& \Longunderstack{Tensor\\ Sea} &\Longunderstack{Without\\Sea}\\\midrule
   \toprule$\vspace{2mm}\Sigma^{*+}$ &0.71084 - 0.17665 r & 0.5347 &0.5127&0.0170&0.0048& 0.5611  &0.5381 &0.0177&0.0050 &4.675&&\\ \vspace{2mm}
       $\Sigma^{*-}$ & -0.35576 - 0.17682r  & -0.5325&-0.5127&-0.0149&-0.0048&-0.5061 &-0.4872&-0.0143&-0.0045&-3.788&&\\ \vspace{2mm}
      $\Sigma^{*0}$ & 0.17585 - 0.17737r & -0.0015&0.0&-0.0015&0.0& 0.0249 & 0.0256&-0.0008&0.0 &-0.177 && \\ \vspace{2mm}
      $\Xi^{*-}$ &-0.17733 - 0.35895r & -0.5362&-0.5082&-0.0123&-0.0157&-0.4827&-0.4576&-0.0108 &-0.0141&-2.359&&\\ \vspace{2mm}
      $\Xi^{*0}$ & 0.36149 - 0.36578r & -0.0042&0.0&-0.0042&0.0 &0.0503 &0.0505&-0.0007&0.0&-0.887&&\\ \vspace{2mm}
      $\Omega^{*-}$ & 0.00007 - 0.52970r & -0.5325&-0.5318&-0.0007&0.0027 & -0.4530&-0.4524&-0.0006 &0.0023& -1.198&&\\ \vspace{2mm} 
      $\Sigma^{+}$ & 0.63812 - 0.062899r & 0.5752&0.4527&0.1227&-0.0003 &0.5846 &0.4470&0.1379 &-0.0003 &3.396&&\\ \vspace{2mm}
      $\Sigma^{-}$ & -0.33284 - 0.033445r & -0.3601 &-0.2001&-0.1664&0.0003&-0.3612&-0.2069&-0.1546 &0.0003 &-2.343&&\\ \vspace{2mm}
      $\Sigma^{0}$ & 0.13594 - 0.18124 r  & 0.0 &-&-&-&-0.0182&-0.0373&0.0190&0.0&-0.1995 &\\ \vspace{2mm}
      $\Xi^{-}$ & -0.03814 - 0.35289r & -0.3910&-0.1953&-0.1960&0.0 &-0.3383&-0.1595&-0.1791 &0.0003& -2.381 &&\\ \vspace{2mm}
      $\Xi^{0}$ & -0.02696 - 0.34976r & -0.3767&-0.4189&0.0421&0.0&-0.3245&-0.3733&
      0.0489 &0.0&-1.090&&\\ \vspace{2mm} 
      $\Lambda^{0}$ & 0.11881 - 0.170227r  & 0.0 &-&-&-&-0.0260&-0.0361&0.0101 &0.0&-0.1994 &&\\  \bottomrule \bottomrule
    \end{tabular}
    }}
     \caption{Numerical values for the charge radii of spin-$(\frac{1}{2})^{+}$ octet and spin-$(\frac{3}{2})^{+}$ decuplet baryons in the units of [fm$^2$] with symmetry breaking parameter r=0.850.}
    \label{tab:3}
\end{table*}
\end{landscape}

\begin{landscape}
\begin{table*}[h!]{\normalsize
\renewcommand{\arraystretch}{1.0}
\tabcolsep 0.1cm
    \centering
   \small{
   \begin{tabular}{cccccc|cccccccc}\toprule\hline
           &           & \multicolumn{8}{c}{Statistical model} &  &\\\cmidrule{3-10}
   Baryons & \Longunderstack{Quadrupole\\ moment} & \Longunderstack{SU(3) Symmetry \\with \\(scalar+vector\\+tensor) sea}& \Longunderstack{Scalar \\Sea}& \Longunderstack{Vector\\ Sea}& \Longunderstack{Tensor\\ Sea} &
   \Longunderstack{SU(3) Symmetry\\ breaking with\\(Valence+sea)} & \Longunderstack{Scalar\\Sea}& \Longunderstack{Vector\\Sea}&\Longunderstack{ Tensor\\ sea} & \Longunderstack{Without\\Sea} \\\midrule
  \toprule \vspace{2mm}$\Sigma^{*+}$ & -0.24520 + 0.06353r & -0.1817&-0.1836&-0.0022& 0.0041 &-0.1911 &-0.1927&-0.0023&0.0039&-0.539&&\\ \vspace{2mm}
   $\Sigma^{*-}$ & 0.12260 + 0.06351r  & 0.1861&0.1836&0.0020&0.0004&0.1766&0.1744 &0.0019&0.0001&0.634&&\\ \vspace{2mm}
   $\Sigma^{*0}$ & -0.06087 + 0.05902r & -0.0018&0.0&-0.0003&-0.0015 &-0.0106&-0.0091 &-0.0002&-0.0012 &-0.036 && \\ \vspace{2mm}
   $\Xi^{*-}$ & 0.06370 + 0.12293r & 0.1866&0.1819&0.0042&0.0004&0.1682&0.1638 &0.0037&0.0006&0.634&& \\ \vspace{2mm}
   $\Xi^{*0}$ & -0.13256 + 0.12805r & -0.0045&0.0&0.003&-0.0048 &-0.0236&-0.0181&0.0 &-0.0005&-0.119&& \\ \vspace{2mm}
   $\Omega^{*-}$ & 0.18816r & 0.1881&0.1904&0.0002&-0.0024&0.1600 &0.1619&0.0 &-0.0021&0.502&& \\ \vspace{2mm}
   $\Sigma^{+}$ & 0.02333 + 0.035296r & 0.0667&0.0012&0.0653&0.0 &0.0533&0.0010&0.0523&0.0 &0.257&&\\ \vspace{2mm}
   $\Sigma^{-}$ & -0.01805 + 0.03079r & 0.0020&0.0011&0.0007&0.0 &0.0081 &0.0009&0.0071&0.0 &0.246&& \\ \vspace{2mm}
   $\Sigma^{0}$ &-0.03602 + 0.02883r  & 0.0 &-&-&-&-0.0114&0.0006&-0.0121 &0.0&-0.1018 && \\ \vspace{2mm}
   $\Xi^{-}$ & 0.03262 - 0.00346r & 0.0362&0.0011&0.0349&0.0001 &0.0296&0.0011&0.0284 &0.0 &0.252& &\\ \vspace{2mm} 
   $\Xi^{0}$ & -0.03152 - 0.01404r& -0.0422&-0.003&-0.038&0.0&-0.0434 &-0.0032&-0.0400 &-0.0001&-0.586&& \\  \vspace{2mm}
   $\Lambda^{0}$ &-0.034183 + 0.031026r  & 0.0 &-&-&-&-0.0077 &0.0007& -0.0084&-0.1018 && \\ \bottomrule \bottomrule
    \end{tabular}
    }}
     \caption{Numerical values for the quadrupole moment of spin-$(\frac{1}{2})^{+}$ octet and spin-$(\frac{3}{2})^{+}$ decuplet baryons in the units of [fm$^2$] with symmetry breaking parameter r=0.850.}
    \label{tab:4}
\end{table*}
\end{landscape}

\end{document}